\documentclass[12pt]{article}
\usepackage{graphicx}
\date{}
\title{\bf
Efimov Trimers : Analytic Solution Via Separable Potentials
  }
\author{
{\normalsize\bf
A.N.Mitra \thanks{Email: (1)ganmitra@nde.vsnl.net.in;
(2)anmitra@physics.du.ac.in}
}\\
\normalsize Dept of Physics, Univ of Delhi, Delhi-110007, India}

\begin{document}

% typeset front matter
\maketitle
\begin{abstract}
The exact dynamical equation for Efimov trimers in the short range
limit of dimer resonances is derived via Yamaguchi separable
potentials. This equation which overcomes the non-uniqueness problem
of zero-range potentials, closely resembles one derived recently by
Gogolin et al, with no further assumptions, and hence enjoys all the
benefits of the latter results.
\end{abstract}

\section{Introduction}

Although Efimov states were proposed nearly four decades ago [1],
they have come into prominence only in recent times following the
experimental detection of a possible candidate in ultra cold atoms
[2]. On the other hand, their mathematical background dates back
even earlier through the work of Danilov [2], on the basis of a
seminal paper by Skoroniakov and Ter-Martirosian [3] for the 3-body
bound states of as many identical bosons, under conditions of near
resonance for dimers with "zero range" forces. In view of the
obvious importance of such states, their theoretical foundations too
have been the subject of renewed scrutiny from different angles. In
particular, Gogolin et al [5](GME), following a method of
Jona-Lasino et al [6], have found an interesting analytical solution
to the spectrum of Efimov states for 3 identical bosons, using the
language of creation and annihilation operators in field theory. In
the process they have also confirmed  Danilov's observation [2] on
the inadequacy of the zero range approximation [3] to produce a
unique Efimov spectrum without additional input (as pointed out by
Gribov [2]), and remedied this defect with a certain parameter
present in their formalism which comes nearest to the more familiar
{\it effective range} theory in nuclear physics.
\par
In this paper we attempt an elementary derivation of the Efimov
effect, in the more orthodox language of nuclear physics via quantum
mechanics, using Yamaguchi separable potentials [7] which had once
been employed for the solution of the bound state 3-body problem
[8]. The reason for this old-fashioned choice is that the system
under study strictly conserves the number of particles (no pair
creation / annihilation effects!) which not only allows one to
dispense with the more advanced language of field theory but also
has a built-in "effective range" feature which automatically
incorporates an additional length dimension on this account (the
$R*$ term of GME [5]), without the need for a fresh insertion by
hand. And since the Yamaguchi separable method was originally
conceived in momentum space for the 3-body problem[8], the {\it
single particle} structure of the GME [5] equations in momentum
space is automatically reproduced without extra efforts. As an extra
bonus, we shall also find a detailed correspondence of the Yamaguchi
formalism [7] to its GME counterpart [5]. In the next section we
recall the basic results of the 2- and 3-body wave functions for
identical particles under pairwise separable potentials [8,9],and
show how they automatically reduce to the GME equations [5], and in
so doing, overcome the non-uniqueness problem of the zero-range
approximation [2,3].

\section{1D Trimers  via Separable Potentials}
\setcounter{equation}{0}
\renewcommand{\theequation}{2.\arabic{equation}}

We now give in barest outline a 3-body formalism [8] with
$separable$ potentials [7], using 3 identical bosons with no
internal d.o.f.s., in a more or less similar notation to the GME
formalism [5]. To that end we start with the essentials of a 2-body
system directly in momentum ($p$) space, as originally employed [8],
where a positive coupling constant $\lambda$ indicates an attractive
interaction, and the units are such that $m=\hbar=1$.

\subsection {The 2-body amplitude on- and off- shell}

 The wave function $\psi$ for a 2-body scattering state in the
c.m. frame satisfies the S.equation [7,8]
\begin{equation}\label{2.1}
(p_1^2 + p_2^2 -E) \psi(q) = \frac {\lambda}{(2\pi)^3}\int d^3q'
g(q)g(q')\psi(q'); \quad E = k^2+i\epsilon;\quad 2q =p_1-p_2
\end{equation}
leading to the off-shell scattering amplitude defined by
$$\psi_k(q)= \delta^3({\bf q-k}) +
\frac{a(q,k)}{(2\pi)^3(q^2-k^2-i\epsilon}$$
which works out as
\begin{eqnarray}\label{2.2}
a(q,k) &=&  \frac{g(p)g(k)}{4\pi(\lambda^{-1}-h(k))} \nonumber \\
h(k)   &=&  \frac{1}{(2\pi)^3}\int d^3q g^2(q)/(q^2 - k^2-i\epsilon)
\end{eqnarray}
where $k^2 > 0$ and the function $g(q)$ is taken as $1/(\beta^2 +
q^2)$ [7]. Then on integration over $q$ the on-shell scattering
amplitude $a(k,k)$=$\sin\delta e^{i\delta}/k$ works out as an
"exact" effective range formula
\begin{equation}\label{2.3}
k \cot\delta =4\pi \lambda^{-1}(\beta^2 + k^2)^2 -\beta/2
+\frac{k^2}{2\beta} \equiv -1/a +\frac{1}{2}r_0 k^2
\end{equation}
from which the set $\lambda;\beta$ is expressible in terms of
$1/a;r_0$ or vice versa. This result is ready for  employment in the
corresponding trimer equations that follow next.

\subsection{3-Body Equation in one-dimensional  form}

The trimer wave function $\Psi$ satisfies an S.equation of the
form[8,9]
\begin{equation}\label{2.4}
D(E)\Psi(q,p) = \sum_i \lambda g(q_i)\int dq_i'\Psi(q_i',p_i)d^3q_i'
\quad D(E)\equiv p_1^2 + p_2^2 + p_3^2 -E
\end{equation}
where $q_1 \equiv (p_2-p_3)/2$, and cyclical permutations.This
equation yields an effectively 2-body structure of the form [8,9]
\begin{equation}\label{2.5}
D(E)\Psi(q,p) \equiv \sum_i g(q_i)\phi(p_i); \quad D(E)= q_i^2 +
3p_i^2/4 - E
\end{equation}
where $\phi$ is a one-dimensional function of the indicated
arguments. On plugging  (2.5) back into the basic form (2.4), one
arrives at an explicit equation for the $\phi$ function :
\begin{equation}\label{2.6}
[\lambda^{-1}- h(k_1)]\phi(p)= "2" \int
d^3p'g(p'+p/2)g(p+p'/2)\phi(p')
\end{equation}
where the integration variable on the rhs has been so adjusted  as
to make the argument of the $\phi$-function equal to the integration
variable, and the factor of $"2"$ in front indicates that the two
"cross" terms give equal contributions. And $h(k_1)$ is the same
function of $k_1=\sqrt{-3p^2/4 + E}$ as $h(k)$ is of $k=\sqrt{E}$ in
eq (2.2), except that for the bound trimer state, $E = -\alpha^2<0$.
\par
Before proceeding further, a few words of comparison of (2.6) with
eq. (4) of GME [5] are in order. The function $\phi(p)$ (which had
been named the "spectator function " in ref [8]), corresponds to the
symbol $\beta_K$ of GME, which may also be called the relative wave
function of a particle of momentum $p$ wrt the pair of the other
two. As to the significance of the various terms in (2.6), the term
$h(k_1)\phi(p)$ on the lhs of our (2.6) has a precise counterpart in
the term $A{K,k}$ of the GME eq(4); and the integral term on the rhs
of (2.6) matches, (complete with the factor of "2"), with the GME
term $ 2 A{k-K/2, -k/2-3K/4}$, where the arguments correspond to
those of our $g$- functions on the rhs of (2.6) $before$ the shifts
in the integration variables were effected. Finally the role of the
length scale $R*$ in GME [5], is automatically incorporated in the
Yamaguchi form factor $g(q)$ [7] inasmuch as the exact effective
range formula (2.3)goes beyond the scattering length formalism of
ref [3] which in turn had led Danilov [2] to the conclusion of
non-uniqueness of that approximation. As we shall see below, Eq (7)
of GME is fully reproduced when (2.6) is subjected to the short
range limit, which also generates a correction term that plays the
role of $R*$ [5].

\section{GME Equation: Result and Conclusion}
\setcounter{equation}{0}
\renewcommand{\theequation}{3.\arabic{equation}}

 To obtain the short range limit of Eq(2.6), one
needs the following steps. First,  Eq(2.3) yields, up to terms of
order $k^2$, the "effective range" relation
\begin{eqnarray}\label{3.1}
\frac{4\pi\beta^4}{\lambda}&=& -\frac{1}{a} - k^2 R_0 +\beta/2
\nonumber \\
-R_0                       &=& r_0/2 -3/2\beta +2/a\beta^2
\end{eqnarray}
which expresses $\lambda^{-1}$ in terms of $a;R_0$. Next, this
expression for$\lambda^{-1}$ must be substituted in (2.6) carefully
enough so that the short range limit $\beta\rightarrow \infty$
emerges smoothly. To that end, the term $\beta/2$ in (3.1) which is
large in this limit, must get cancelled by a corresponding term in
the function $h(k_1)$, so that it makes sense to combine these terms
together before other operations are performed. The result of this
step is to produce a combination which yields simply
\begin{equation}\label{3.2}
\beta/2 - 4\pi\beta^4 h(k_1) = \int\frac{d^3\beta^4
g^2(q)(3p^2/4+\alpha^2} {(2\pi^2 q^2(q^2+3p^2/4)}\rightarrow
\sqrt{3p^2/4+\alpha^2}
\end{equation}
the last step being taken in the short range limit. The rest is
straightforward: Just multiply both sides of (2.6) by $4\pi\beta^4$
after substituting for $\lambda^{-1}$ from (3.1), and take the limit
of large $\beta$, noting that $\beta^2\times g$ approaches unity.
The final result is
\begin{equation}\label{3.3}
[-1/a +R_0(3p^2/4+\alpha^2)+\sqrt{3p^2/4+\alpha^2}]\phi(p)=\pi^{-2}
\int\frac{d^3p'\phi(p')}{p^2+p'^2+p.p'}
\end{equation}
which on azimuthal integration and a transformation
$p\phi(p)=\psi(p)$ yields the GME equation (7) [5], with the
identification $R_0 \Leftrightarrow R*$.

To conclude, we have obtained a 1D equation via Yamaguchi separable
potentials which in the short range limit closely resembles the GME
[5] equation under identical conditions for dimer resonances. Since
there  already exists a vast literature on this subject, including
educational ones [10], we need hardly comment further, except to
claim that all the benefits of the GME results,subsequent to their
eq (7), also accrue to the present formalism.
\par
I am grateful to Indranil Majumdar for bringing the recent
developments on the Efimov effect to my notice.

\end{document}